\documentclass[sigconf]{acmart}

\usepackage[ruled]{algorithm2e}
\usepackage{stmaryrd}
\usepackage{graphicx}
\usepackage{color}
\usepackage{comment}
\usepackage{url}
\usepackage{wrapfig}
\usepackage{enumerate}
\usepackage{epstopdf}
\begin{document}

\acmYear{2018} 
\setcopyright{acmcopyright}
\acmConference[MMSys'18]{ACM Multimedia Systems 2018}{June 2018}{Amsterdam, Netherlands}
\acmPrice{15.00}
\acmDOI{http://dx.doi.org/10.1145/3083187.xxxx}
\acmISBN{978-1-4503-xxxx-x/xx/xx}

\renewcommand{\baselinestretch}{1.5}

\title{Deep Live Video Ad Placement on the 5G Edge}

\author{Mohammad Hosseini}
\affiliation{\institution{University of Illinois at Urbana-Champaign (UIUC)}}
\email{shossen2@illinois.edu}


\begin{abstract}
The video broadcasting industry has been growing significantly in the recent years, specially on delivering personalized contents to the end users. While video broadcasting has continued to grow beyond TV, video adverting has become a key marketing tool to deliver targeted messages directly to the audience. However, unfortunately for broadband TV, a key problem is that the TV commercials target the broad audience, therefore lacking user-specific and personalized ad contents.

In this paper, we propose a deep edge-cloud ad-placement system, and briefly describe our methodologies and the architecture of our designed ad placement system for delivering both the Video on Demand (VoD) and live broadcast TV contents over MMT streaming protocol. The aim of our paper is to showcase how to enable targeted, personalized, and user-specific advertising services deployed on the future 5G MEC platforms, which in turn can have high potentials to increase ad revenues for the mobile operator industry.

\end{abstract}



\maketitle
The video streaming industry has witnessed significant growth in the recent years, specially on delivering personalized contents to the end users. While videos have continued to grow beyond TV, video advertising have become a key business tool to deliver targeted messages directly to a wide audience. Statistics show that the US TV commercials produced 70.6 billion dollars in revenues in 2016 alone \cite{tvad1, dashhosseini, hosseini2014}, and more specifically, 88\% of short ads (of around 30 seconds), are watched on mobile devices \cite{tvad2}. Unfortunately, for broadband TV, a key problem is that the commercials target the broad audience and lack user-specific and personalized ad contents.

At the same time, use of AI-driven approaches on the cloud especially leveraging deep learning has become a \textit{de facto} to provide computing-intensive processing services and automated solutions. With the rapid growth of video streaming and over-the-top (OTT) streaming market, leveraging AI can be a key marketing tool to personalize user experience and recommendations, for instance to deliver targeted video commercials.

In this work, we showcase a deep edge-cloud ad-placement prototype, and briefly discuss how we designed our system for both VoD services (offline mode) and live streaming from a broadcast TV. We employ the emerging MPEG Media Transport (MMT) protocol, and use that as our underlying streaming protocol. We implemented our ad detection system using both conventional cross-correlation based approaches as well as deep neural network methodology. The aim of our ad-placement prototype is to envision a targeted advertising service on the 5G cloud edge to replace general ad contents with more personalized and user-specific commercial contents. We implemented our system in response to the emergence of the 5G MEC and MPEG Network-based Media Processing (MPEG NBMP) initiatives, and showcase how it can be leveraged to provide personalized mobile services on the edge, potentially to increase advertising revenues for the mobile operator industry.

The paper is organized as follows. In Section \ref{background}, we briefly cover some background information and related work. In Section \ref{design}, we explain the design of our deep ad-placement prototype, and discuss the architecture and how we implemented the system components. We finally conclude the paper in Section \ref{conclusion}.

\section{Background}
\label{background}

\begin{figure}[!t]
\centering
\includegraphics[width=.8\columnwidth]{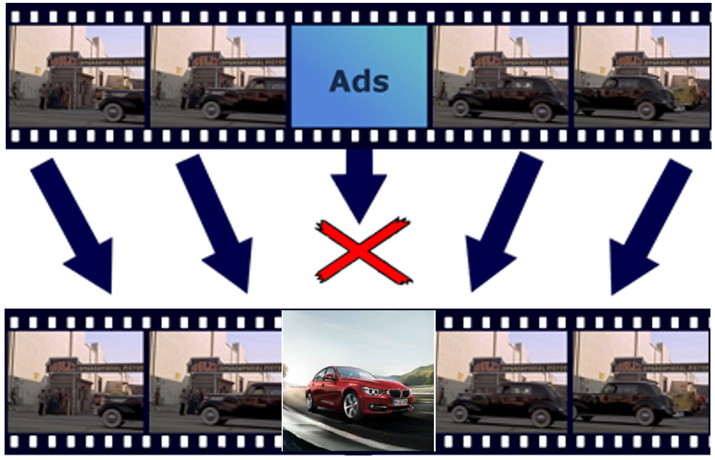}
\caption{An example illustration of our ad placement system.}
\label{example}
\end{figure}
\subsection{MMT Protocol}

\begin{figure*}[!t]
\centering
\includegraphics[width=.9\textwidth]{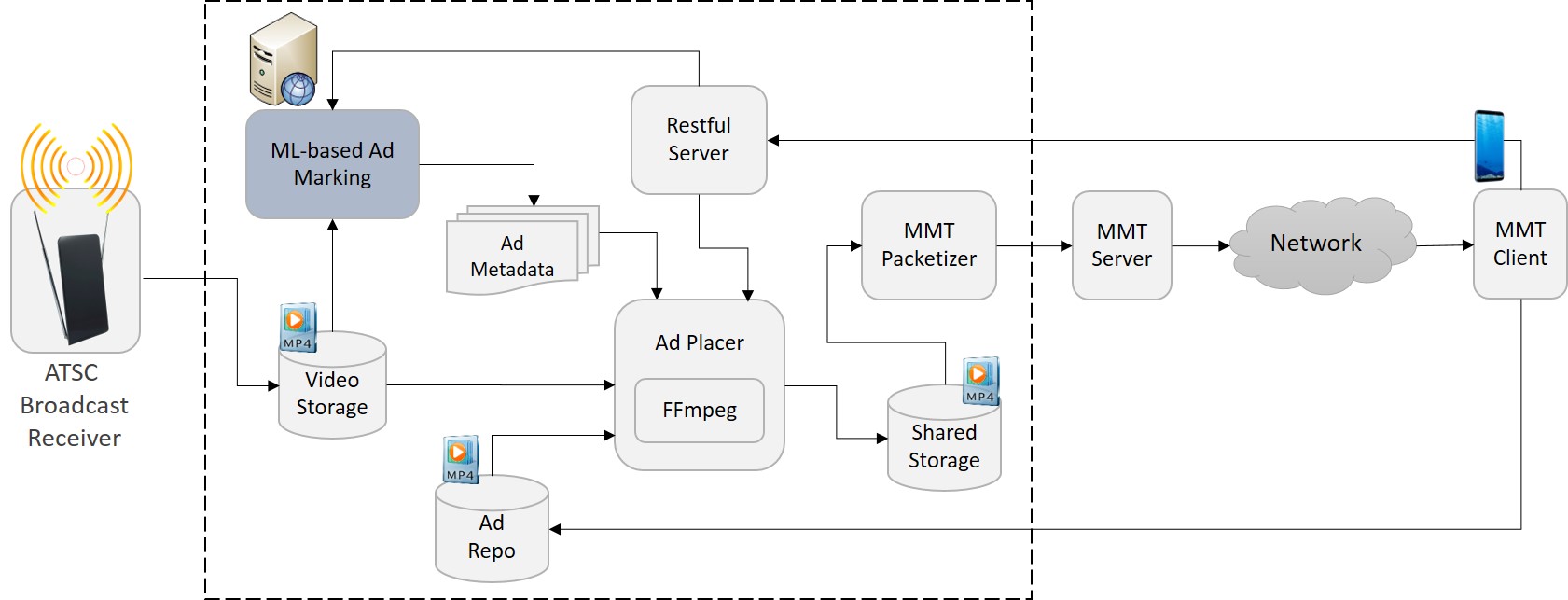}
\caption{Abstract structure of our ad-replacement edge cloud system.}
\label{adplacement}
\end{figure*}
Traditional one-way multimedia streaming services such as TV broadcasting using MPEG-TS or media delivery technologies have had the necessity to adapt with the trending MPEG multimedia technologies. The driving technologies for MMT emergence specifically includes IP convergence, dynamic and flexible multimedia consumption, as well as the Information-Centric Networks (ICN) \cite{mmt1}.

MPEG Media Transport (MMT), as specified by ISO/IEC 23008-13 \cite{mmt2}, is an emerging streaming protocol which provides a functionality for the delivery of coded media data for multimedia services over concatenated heterogeneous all-IP networks. All-IP emergence has put a pressure on the TV broadcast services delivered over the air or via cable networks to adopt IP for the future broadcast TV networks. In addition, the increasing demand of more personalized services made end users more willing to consume media contents matching their preferences. Therefore, supporting dynamic and flexible multimedia contents transmission and consumption has become a key design element of MMT protocol. Due to MMT's compliance with content personalization over an all-IP broadcast TV, we employ MMT as the underlying streaming protocol inside our system.

\subsection{Cloud Media Processing and MPEG NBMP}
Cloud computing aims to provide a distributed platform to perform various resource-intensive processing tasks in the network. Recently, many AI-driven cloud applications such as recommendation systems \cite{recommendation}, autonomous cars \cite{cars}, and live video analytics \cite{msr} have been studied on feasibility of deploying deep learning-based functionalities on the cloud edge for security, privacy, as well as latency improvements. There are also several industry practices such as \cite{google}, \cite{azure}, \cite{alibaba}, and \cite{bitmovin} to name a few, to offer cloud media processing solutions, including live media encryption, adaptive streaming, or basic AI analytic such as object or scene change detection.

On the MPEG side, a new ad-hoc group has been initiated for Network-based Media Processing, or in short, MPEG NBMP \cite{nbmp}. Large-scale resource-intensive media processing tasks deemed to exhaust the client resources are offloaded to the cloud to benefit from the resource-extensive cloud platform provided by NBMP, also to respect the client's limited resources. Our proof-of-concept demonstrated here is implemented in response to the MPEG NBMP functionality and requirements.

\section{System Design and Structure}
\label{design}
The aim of our work is to showcase a system packaged as a container mounted on a edge-cloud server, as enabled by 5G MEC, aimed to benefit the operator's services with lower service latency and enhanced user awareness. In our prototype, an automated ad-replacement system is developed where ad video portions of a linear video stream are detected through AI engines, and are replaced with a secondary ad content supposedly a more personalized and user-specific ad. Figure \ref{example} illustrates the concept. A video content, possibly containing temporal commercial ads, is given as an input to our cloud media processing platform. Our developed system can support VoD scenarios as well as live scenarios where broadcast TV streams are continuously ingested to our cloud system. AI engines residing on the cloud edge recognize the ad contents inside the input video contents. Upon processing, the detected ad contents are then replaced with secondary ad videos, based on the user's location, view history, or other preferences. The output video streams are then aggregated together, and the new video is streamed to the client via MMT protocol.

Figure \ref{adplacement} illustrates an abstract architecture of our AI-driven cloud ad-replacement system, with the dashed area identifying components inside the virtualization core, packaged as a LXC container deployed on a hexa-core Intel Xeon Core i7-6900K running 64-bit Ubuntu 17.04 LTS. In the following, we enumerate the major components of our system.

\subsection{Broadcast TV Receiver}
\begin{figure}[!t]
\centering
\includegraphics[width=.4\columnwidth]{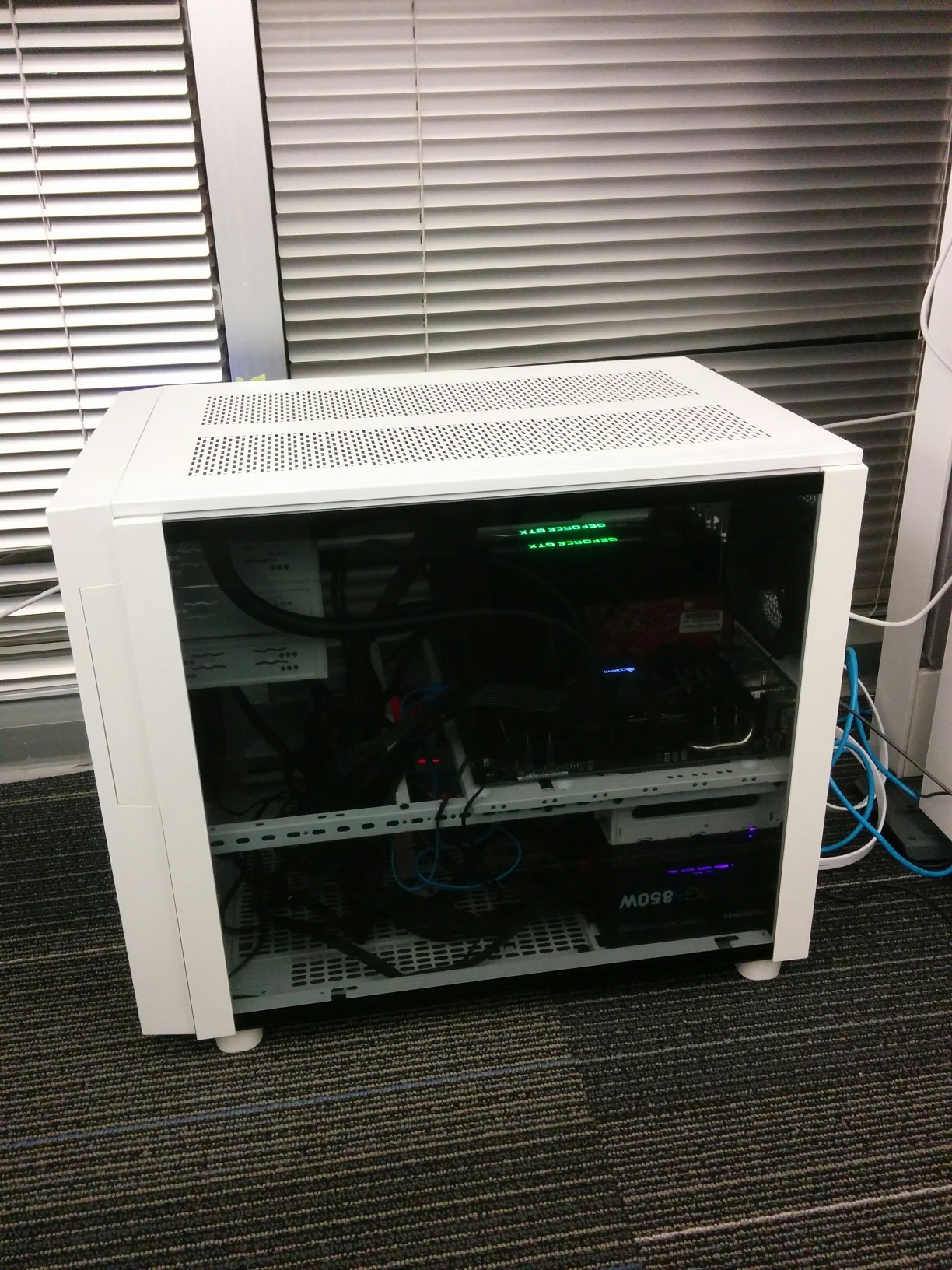}~~~\includegraphics[width=.4\columnwidth]{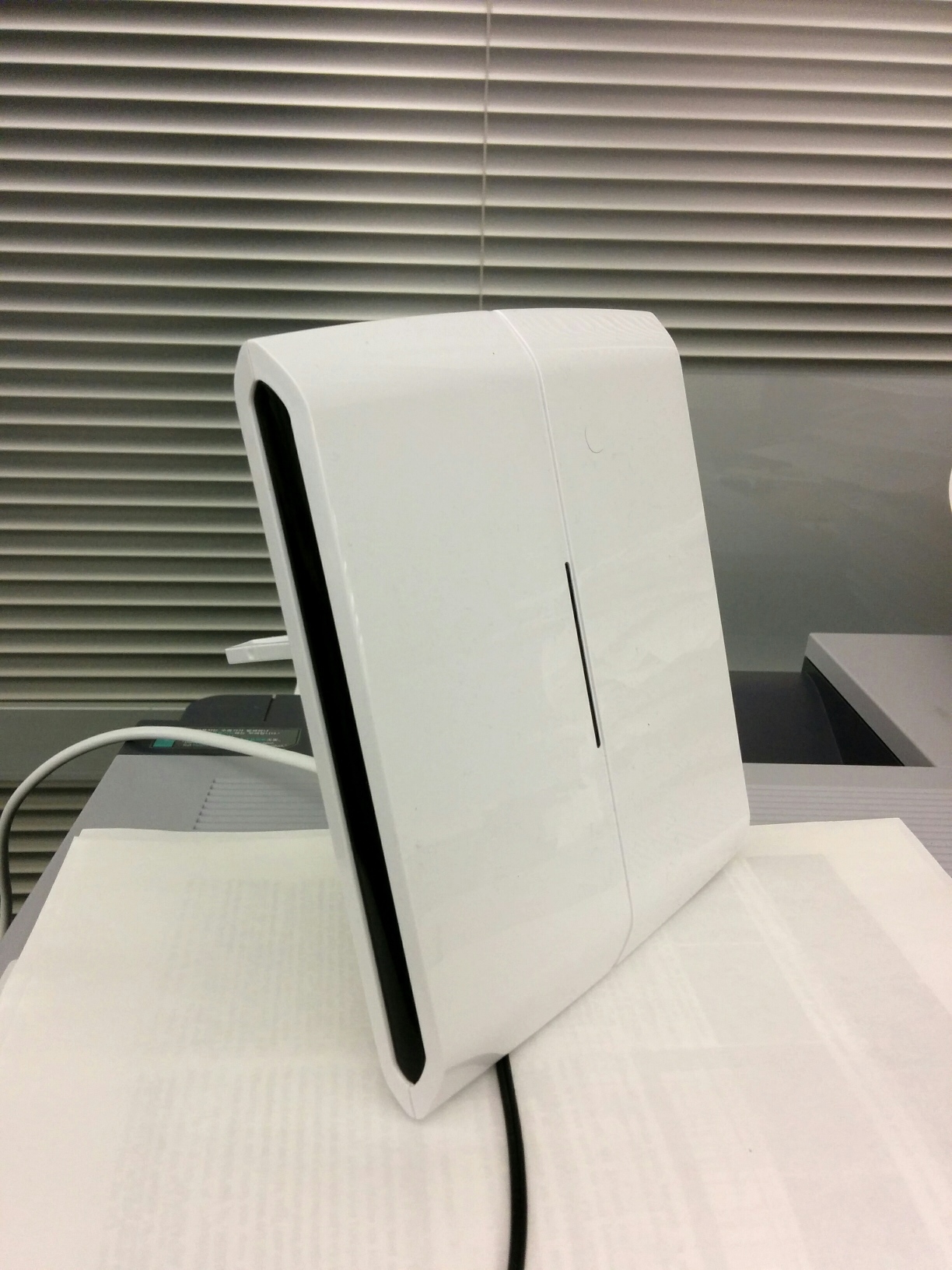}
\caption{Our server and the ATSC digital TV receiver setup.}
\label{server}
\end{figure}

To enable live media processing, we used August DTA300W, an ATSC digital TV antenna with 20dB gain amplifier integrated with our server machine that generates audio and video signals that are picked up from over-the-air broadcast TV. We used the open-source Tvheadend 4.2 TV streaming server \cite{tvheadend}, and configured the ATSC receiver to feed live media contents from a tuned TV channel, in our case, Laff\cite{laff} and Bounce\cite{bounce} TV channels. We implemented an x264-based transcoder to ingest the MPEG-TS streams, temporally segments the live streams into videos of shorter duration while encoding and packaging, and records the audio and video files in a shared storage for processing.

Figure \ref{server} demonstrates our server machine together with the ATSC digital TV receiver setup.

\subsection{RESTful Server}
Given the distributed nature of our prototype, a RESTful server residing on the edge cloud is designed to remotely communicate with the user, receive the client's media processing requests, and initiate the specified media processing pipeline. A client initiates the ad-placement media processing service through communicating with the RESTful server and sending a request through the RESTful API that we have developed. We used Node.js 6.11 with Express 4.15.3 web application framework to develop our REST server and API.

For convenient deployment, we implemented some utility functionality on the server to notify the client with updates on the media processing tasks, such as the user's geographical information, server's specs, and the total processing time. The RESTful server then prepares the media processing pipeline, deploys the media processing container, and starts the ad detection AI engine. The source stream, either the source video uploaded by the client in the VoD mode or the broadcast TV streams in the live mode, are referenced through a unique URI. After that, the AI engine processes the source video stream, and identifies the frame sequence of ad contents in the source video stream.

\subsection{AI Engine}

\begin{figure}[!t]
\centering
\includegraphics[width=.9\columnwidth]{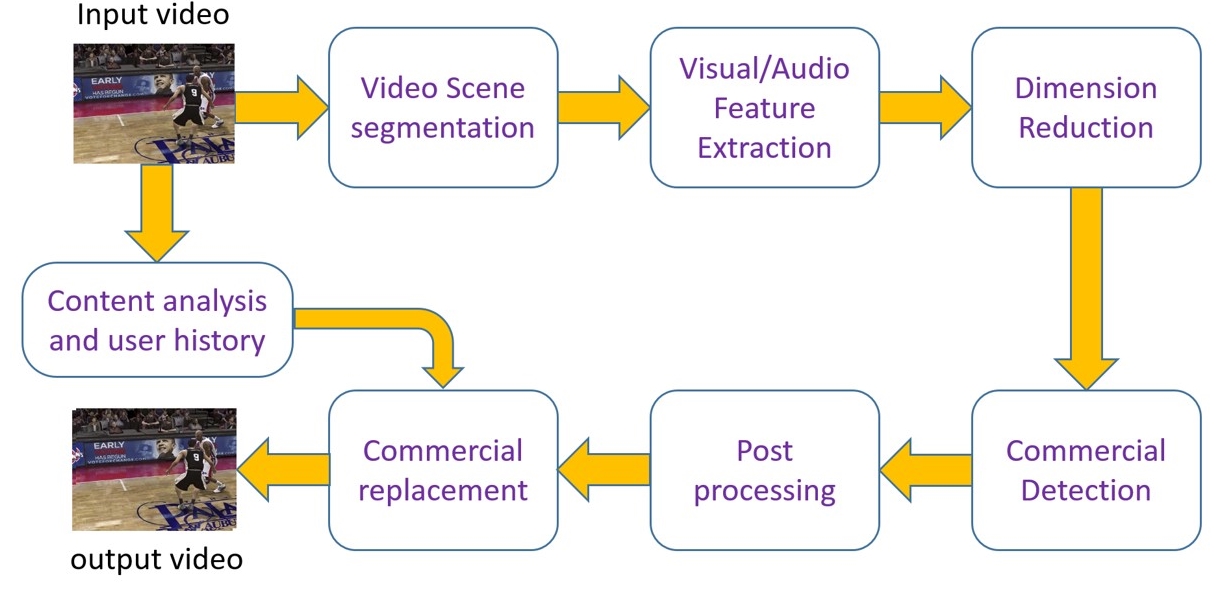}
\caption{The workflow of our deep neural network AI engine.}
\label{ml}
\end{figure}

The main purpose of the AI engine is ad detection and content categorization in a given video segment. To achieve that, we proposed a novel deep neural network that extracts certain features using the audio-visual information of a video content, and performs ad detection and content recognition simultaneously. We developed our deep neural network using TensorFlow 1.3 with Python 3.5 packaged in Anaconda 4.2.

\begin{figure}[!t]
\centering
\includegraphics[width=.85\columnwidth]{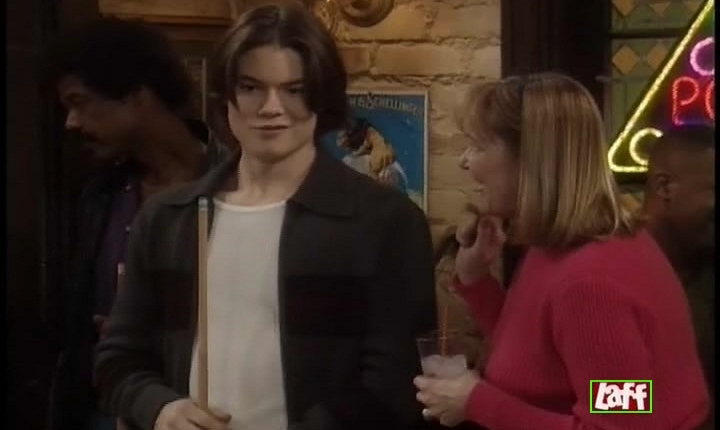}
\caption{Our cross-correlation AI engine as applied on an example frame from Laff TV.}
\label{laff}
\end{figure}

Figure \ref{ml} briefly illustrates the deep-learning workflow in use by our AI engine. First, a shot detection and segmentation module splits a given video segment into multiple shots, each containing a consistent scene. The process is done through iteratively performing a frame difference analysis on consecutive frames to derive the probability distribution of a scene cut. Next, we develop a deep neural network to extract a shot's audio-visual features, providing a basis to distinguish an ad content from a general non-ad video content. To achieve that, we perform a Mel-Frequency Cepstral Coefficients (MFCC) feature extraction method followed by dimension reduction and normalization. The list of features are then evaluated to determine the subset of features that must be used to classify the content. Once the optimum feature subset is selected, they are fused with post-processing to use for classification, and a target ad from a suitable category is identified for replacing the original ad content. We use \%80 of a test dataset as training, and the rest for testing.

When extending our prototype to live contents, we witnessed that all the broadcast TV networks place the TV logo on the movie contents, while the logo is taken off when commercial contents are shown. To enable ad content detection based on that cue, we developed a simple AI module based on cross-correlation approach, which selectively analyzes any linear video streams, and finds a specific pattern in a long-duration signal, in this case, a single frame within a video segment. When a TV network logo is present in a video frame, our AI algorithm generates a vigorous cross-correlation peak at its approximate center compared to a case where no logo is present. We then apply a defined threshold to detect this peak, leading to detection of the logo presence and therefore, a correct classification. We used OpenCV 3.3 to implemented our logo detection AI engine.

Figure \ref{laff} shows the visual view of how our cross-correlation AI engine works on an example frame from Laff broadcast TV. Our cross-correlation algorithm is light-weight, high precision, doesn't require re-training, and runs significantly faster compared to our deep-learning based approach on processing of live broadcast TV contents. Table \ref{tab:comparison} provides a quick comparison of the two AI engines on the broadcast content processing.

\begin{table}[!h]
  \begin{center}
    \caption{Comparison of the two AI engines on broadcast TV.}
    \label{tab:comparison}
    \begin{tabular}{l|c|c}
    
      \textbf{ } & \textbf{Execution Time (ms)} & \textbf{Accuracy}\\
      \hline
      Deep-Learning & 7477 ms & \%80.2 \\
      \hline
      Cross Correlation & 83 ms & $\approx$ \%100 \\
      \hline
    \end{tabular}
  \end{center}
  \vspace{-.3cm}
\end{table}

\subsection{Ad Meta-Data}
The output of the AI engines is a text meta-data, which mainly includes time intervals of ad contents within the source video stream and a target ad content ID. In specific, the generated ad meta-data includes the following fields:

\begin{itemize}
\item ad content's start timestamp
\item ad content's end timestamp
\item ad content's start frame number
\item ad content's end frame number
\item is\underline{ }ad flag indicating if the content specified is ad (optional- always set to 1 for ads).
\item target\underline{ }ad\underline{ }ID specifying the URI of the target ad that is going to replace the original ad content.
\end{itemize}

\subsection{Ad Placer}
After the ad meta-data is generated, it is then passed to an ad-placer module which substitutes the original ad with a new ad content given the target ad ID, and performs a concatenation on the whole sub-streams. We used Java 1.8 wrapped around FFmpeg 3.3.6 (Hilbert) as the core engine of our video splitting and stitching processes. The ad-placer module is interfaced with a shared storage to retrieve the source video segments, an ad repository for the target ads, as well as a shared storage to place the generated output videos. Upon completion of the ad-placement process, the generated video stream is then passed to the MMT streaming server, where the media contents are packetized and prepared for transmission. The generated MMT packets are then transferred to an MMT delivery server, which pushes the MMT packets to an MMT-compliant client for playout.

\subsection{MMT Streaming Server}
We implemented our streaming server on top of the MMT reference software package as a part of the MPEG-H standard in ISO/IEC 23008-4 \cite{mmtrefsw}. The MMT streaming server consists of two components: The MMT packetizer, and MMT transmission server. The MMT packetizer takes a configuration file as an input with the location of input assets defined, reads fragmented x264 MP4-formatted audio and video files as its inputs, and encapsulates them into MMT Media Processing Units (MPU). The packetizer then generates the MMT flow as a multiplex of numbered packets, and stores them into a pre-configured shared location prepared for transmission. The role of MMT transmission server is then to take the MMT packets and transmit them to a registered MMT client using WebSocket protocol.

\begin{figure}[!t]
\centering
\includegraphics[width=.9\columnwidth]{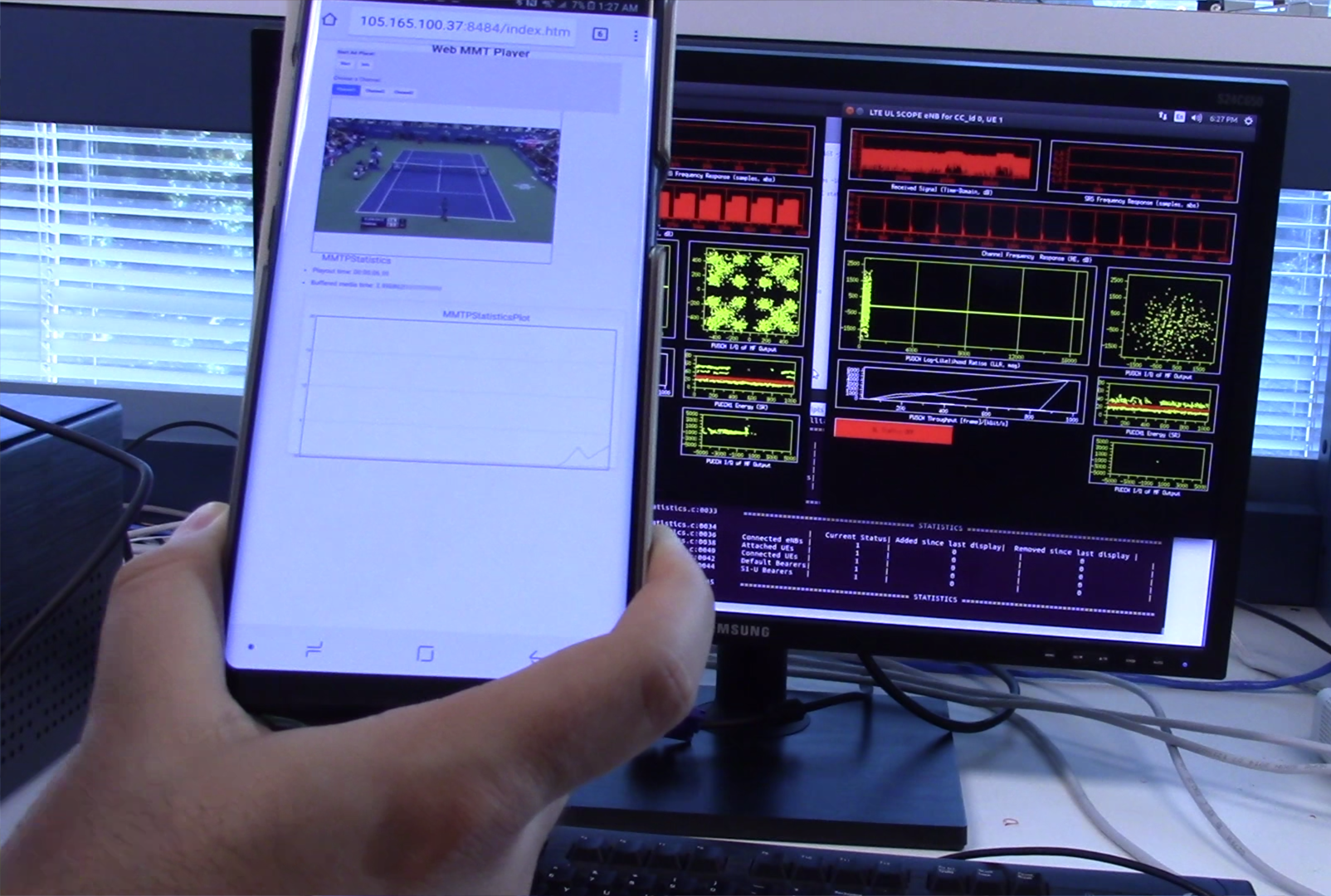}
\caption{An example MMT player running on a Galaxy S8 client connected to our 5G MEC emulator.}
\label{client}
\end{figure}

\subsection{MMT Client}
The MMT client receives an MMT flow, buffers them, de-multiplexes them, and then passes the packets to a reconstruction module to extract the payload and reconstruct the MPU. The MPUs are then sent to a decoding engine for playback. We implemented the MMT player using web-based technologies, making it suitable for a heterogeneous exposure via a JavaScript interface in HTML 5 compliant browsers. The interface provides a full-duplex communication channel, which also enables interactive functionality to initiate cloud media processing tasks and retrieve additional information such as the processing server's specs, the total processing time, and other information. We also implemented a statistics module on the client that plots some of the players information, specifically the total playback time elapsed and the total time of buffered media. 

Figure \ref{client} illustrates an example playback session running on a Galaxy S8 smartphone, with the MMT-encapsulated media streamed over a 5G MEC testbed that we have developed. A short video of our prototype for VoD mode is also provided.

\section{Conclusion}
\label{conclusion}
In this paper, we demonstrate and explain the design of an AI-driven ad-placement prototype placed on the edge cloud, and discuss how we designed our system for both VoD and live broadcast TV contents. We implemented our ad detection system using both cross-correlation as well as deep neural network-based approaches, and used the emerging MPEG Media Transport (MMT) as our streaming protocol. Our AI-based ad-placement prototype is aimed to help content providers and mobile operators to deliver targeted advertisement services with more personalized and user-specific ads on the cloud edge closer to the end users, possibly through the features enabled by upcoming 5G MEC and MPEG NBMP platforms.

\section{Acknowledgments}
Research reported in this paper was conducted in collaboration with Samsung. We would also like to thank Prakash Kolan and Shervin Minaee for their assistance throughout different aspects of this study.

\bibliographystyle{ACM-Reference-Format}
\bibliography{sigproc}


\begin{thebibliography}{00}


\ifx \showCODEN    \undefined \def \showCODEN     #1{\unskip}     \fi
\ifx \showDOI      \undefined \def \showDOI       #1{{\tt DOI:}\penalty0{#1}\ }
  \fi
\ifx \showISBNx    \undefined \def \showISBNx     #1{\unskip}     \fi
\ifx \showISBNxiii \undefined \def \showISBNxiii  #1{\unskip}     \fi
\ifx \showISSN     \undefined \def \showISSN      #1{\unskip}     \fi
\ifx \showLCCN     \undefined \def \showLCCN      #1{\unskip}     \fi
\ifx \shownote     \undefined \def \shownote      #1{#1}          \fi
\ifx \showarticletitle \undefined \def \showarticletitle #1{#1}   \fi
\ifx \showURL      \undefined \def \showURL       #1{#1}          \fi
\providecommand\bibfield[2]{#2}
\providecommand\bibinfo[2]{#2}
\providecommand\natexlab[1]{#1}
\providecommand\showeprint[2][]{arXiv:#2}

\bibitem[\protect\citeauthoryear{??}{mmt}{2012}]%
        {mmtrefsw}
 \bibinfo{year}{2012}\natexlab{}.
\newblock \bibinfo{title}{ISO/IEC CD 23008-1 MPEG Media Transport}.
\newblock
  \bibinfo{howpublished}{https://mpeg.chiariglione.org/standards/mpeg-h/mpeg-media-transport}.
    (\bibinfo{year}{2012}).
\newblock


\bibitem[\protect\citeauthoryear{??}{ali}{2017}]%
        {alibaba}
 \bibinfo{year}{2017}\natexlab{}.
\newblock \bibinfo{title}{Alibaba Cloud Multimedia}.
\newblock
  \bibinfo{howpublished}{https://alibabacloud.com/solutions/multimedia}.
  (\bibinfo{year}{2017}).
\newblock


\bibitem[\protect\citeauthoryear{??}{bit}{2017}]%
        {bitmovin}
 \bibinfo{year}{2017}\natexlab{}.
\newblock \bibinfo{title}{Bitmovin Video Infrastructure}.
\newblock \bibinfo{howpublished}{https://bitmovin.com/}.
  (\bibinfo{year}{2017}).
\newblock


\bibitem[\protect\citeauthoryear{??}{bou}{2017}]%
        {bounce}
 \bibinfo{year}{2017}\natexlab{}.
\newblock \bibinfo{title}{Bounce TV network}.
\newblock \bibinfo{howpublished}{http://www.bouncetv.com}.
  (\bibinfo{year}{2017}).
\newblock


\bibitem[\protect\citeauthoryear{??}{goo}{2017}]%
        {google}
 \bibinfo{year}{2017}\natexlab{}.
\newblock \bibinfo{title}{Google Cloud Platform}.
\newblock \bibinfo{howpublished}{https://cloud.google.com}.
  (\bibinfo{year}{2017}).
\newblock


\bibitem[\protect\citeauthoryear{??}{laf}{2017}]%
        {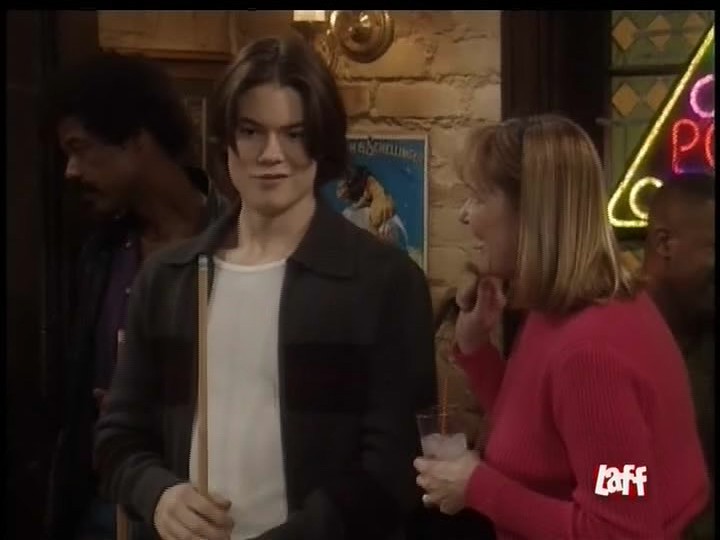}
 \bibinfo{year}{2017}\natexlab{}.
\newblock \bibinfo{title}{Laff Media TV network}.
\newblock \bibinfo{howpublished}{http://www.laff.com}.
  (\bibinfo{year}{2017}).
\newblock


\bibitem[\protect\citeauthoryear{??}{azu}{2017}]%
        {azure}
 \bibinfo{year}{2017}\natexlab{}.
\newblock \bibinfo{title}{Microsoft Azure}.
\newblock
  \bibinfo{howpublished}{https://azure.microsoft.com/en-us/services/media-services}.
    (\bibinfo{year}{2017}).
\newblock


\bibitem[\protect\citeauthoryear{??}{nbm}{2017}]%
        {nbmp}
 \bibinfo{year}{2017}\natexlab{}.
\newblock \bibinfo{title}{MPEG NBMP}.
\newblock
  \bibinfo{howpublished}{https://mpeg.chiariglione.org/standards/exploration/network-based-media-processing}.
    (\bibinfo{year}{2017}).
\newblock


\bibitem[\protect\citeauthoryear{??}{tvh}{2017}]%
        {tvheadend}
 \bibinfo{year}{2017}\natexlab{}.
\newblock \bibinfo{title}{Tvheadend TV streaming server for Linux}.
\newblock \bibinfo{howpublished}{https://tvheadend.org}.
  (\bibinfo{year}{2017}).
\newblock


\bibitem[\protect\citeauthoryear{??}{tva}{2017b}]%
        {tvad1}
 \bibinfo{year}{June 2017}\natexlab{b}.
\newblock \bibinfo{title}{TV advertising revenue in the US 2016-2021}.
\newblock
  \bibinfo{howpublished}{https://www.statista.com/statistics/259974/tv-advertising-revenue-in-the-us}.
    (\bibinfo{year}{June 2017}).
\newblock


\bibitem[\protect\citeauthoryear{??}{tva}{2017a}]%
        {tvad2}
 \bibinfo{year}{May 2017}\natexlab{a}.
\newblock \bibinfo{title}{Social Business: Why Do You Need to Use Social Media
  Videos?}
\newblock
  \bibinfo{howpublished}{https://socialnewsdaily.com/66157/why-do-you-need-to-include-social-media-videos}.
    (\bibinfo{year}{May 2017}).
\newblock


\bibitem[\protect\citeauthoryear{??}{mmt}{2017}]%
        {mmt2}
 \bibinfo{year}{October 2017}\natexlab{}.
\newblock \bibinfo{title}{WD of ISO/IEC 23008-13:2017, Part 13: MPEG Media
  Transport Implementation Guidelines}.
\newblock \bibinfo{howpublished}{https://www.iso.org/standard/70438.html}.
  (\bibinfo{year}{October 2017}).
\newblock


\bibitem[\protect\citeauthoryear{Hosseini, Jiang, Berlin, Sha, and
  Song}{Hosseini et~al\mbox{.}}{2017}]%
        {dashhosseini}
\bibfield{author}{\bibinfo{person}{Mohammad Hosseini}, \bibinfo{person}{Yu
  Jiang}, \bibinfo{person}{Richard~R. Berlin}, \bibinfo{person}{Lui Sha}, {and}
  \bibinfo{person}{Houbing Song}.} \bibinfo{year}{2017}\natexlab{}.
\newblock \showarticletitle{Toward Physiology-Aware DASH: Bandwidth-Compliant
  Prioritized Clinical Multimedia Communication in Ambulances}.
\newblock \bibinfo{journal}{{\em IEEE Transactions on Multimedia\/}}
  \bibinfo{volume}{19}, \bibinfo{number}{10} (\bibinfo{year}{2017}),
  \bibinfo{pages}{2307--2321}.
\newblock
\showDOI{%
\url{http://dx.doi.org/10.1109/TMM.2017.2733298}}


\bibitem[\protect\citeauthoryear{Hosseini, Peters, and Shirmohammadi}{Hosseini
  et~al\mbox{.}}{2014}]%
        {hosseini2014}
\bibfield{author}{\bibinfo{person}{Mohammad Hosseini}, \bibinfo{person}{Joseph
  Peters}, {and} \bibinfo{person}{Shervin Shirmohammadi}.}
  \bibinfo{year}{2014}\natexlab{}.
\newblock \showarticletitle{Energy-Efficient 3D Texture Streaming for Mobile
  Games}. In \bibinfo{booktitle}{{\em Proceedings of Workshop on Mobile Video
  Delivery}} {\em (\bibinfo{series}{MoViD'14})}.
  \bibinfo{publisher}{Association for Computing Machinery},
  \bibinfo{address}{New York, NY, USA}, \bibinfo{pages}{1–6}.
\newblock
\showISBNx{9781450327077}
\showDOI{%
\url{http://dx.doi.org/10.1145/2579465.2579471}}


\bibitem[\protect\citeauthoryear{Kehoe, Patil, Abbeel, and Goldberg}{Kehoe
  et~al\mbox{.}}{2015}]%
        {cars}
\bibfield{author}{\bibinfo{person}{B. Kehoe}, \bibinfo{person}{S. Patil},
  \bibinfo{person}{P. Abbeel}, {and} \bibinfo{person}{K. Goldberg}.}
  \bibinfo{year}{2015}\natexlab{}.
\newblock \showarticletitle{A Survey of Research on Cloud Robotics and
  Automation}.
\newblock \bibinfo{journal}{{\em IEEE Transactions on Automation Science and
  Engineering\/}} \bibinfo{volume}{12}, \bibinfo{number}{2}
  (\bibinfo{date}{April} \bibinfo{year}{2015}), \bibinfo{pages}{398--409}.
\newblock
\showISSN{1545-5955}


\bibitem[\protect\citeauthoryear{Lim, Park, Lee, Aoki, and Fernando}{Lim
  et~al\mbox{.}}{2013}]%
        {mmt1}
\bibfield{author}{\bibinfo{person}{Y. Lim}, \bibinfo{person}{K. Park},
  \bibinfo{person}{J.~Y. Lee}, \bibinfo{person}{S. Aoki}, {and}
  \bibinfo{person}{G. Fernando}.} \bibinfo{year}{2013}\natexlab{}.
\newblock \showarticletitle{MMT: An Emerging MPEG Standard for Multimedia
  Delivery over the Internet}.
\newblock \bibinfo{journal}{{\em IEEE MultiMedia\/}} \bibinfo{volume}{20},
  \bibinfo{number}{1} (\bibinfo{date}{Jan} \bibinfo{year}{2013}),
  \bibinfo{pages}{80--85}.
\newblock
\showISSN{1070-986X}


\bibitem[\protect\citeauthoryear{McMahan, Moore, Ramage, Hampson, and
  y~Arcas}{McMahan et~al\mbox{.}}{2017}]%
        {recommendation}
\bibfield{author}{\bibinfo{person}{Brendan McMahan}, \bibinfo{person}{Eider
  Moore}, \bibinfo{person}{Daniel Ramage}, \bibinfo{person}{Seth Hampson},
  {and} \bibinfo{person}{Blaise~Aguera y Arcas}.}
  \bibinfo{year}{2017}\natexlab{}.
\newblock \showarticletitle{{Communication-Efficient Learning of Deep Networks
  from Decentralized Data}}. In \bibinfo{booktitle}{{\em Proceedings of the
  20th International Conference on Artificial Intelligence and Statistics}}
  {\em (\bibinfo{series}{Proceedings of Machine Learning Research})},
  \bibfield{editor}{\bibinfo{person}{Aarti Singh} {and} \bibinfo{person}{Jerry
  Zhu}} (Eds.), Vol.~\bibinfo{volume}{54}. \bibinfo{publisher}{PMLR},
  \bibinfo{address}{Fort Lauderdale, FL, USA}, \bibinfo{pages}{1273--1282}.
\newblock


\bibitem[\protect\citeauthoryear{Zhang, Ananthanarayanan, Bodik, Philipose,
  Bahl, and Freedman}{Zhang et~al\mbox{.}}{2017}]%
        {msr}
\bibfield{author}{\bibinfo{person}{Haoyu Zhang}, \bibinfo{person}{Ganesh
  Ananthanarayanan}, \bibinfo{person}{Peter Bodik}, \bibinfo{person}{Matthai
  Philipose}, \bibinfo{person}{Paramvir Bahl}, {and}
  \bibinfo{person}{Michael~J. Freedman}.} \bibinfo{year}{2017}\natexlab{}.
\newblock \showarticletitle{Live Video Analytics at Scale with Approximation
  and Delay-Tolerance}. In \bibinfo{booktitle}{{\em 14th {USENIX} Symposium on
  Networked Systems Design and Implementation ({NSDI} 17)}}.
  \bibinfo{publisher}{{USENIX} Association}, \bibinfo{address}{Boston, MA},
  \bibinfo{pages}{377--392}.
\newblock
\showISBNx{978-1-931971-37-9}


\end{thebibliography}

\end{document}